\documentclass[useAMS,usenatbib]{mn2e}
\usepackage{amsmath,amssymb,epsfig,natbib,bm,psfrag,ifthen,url}
\usepackage{hyperref}
\usepackage{epsfig}
\usepackage{graphicx}
\usepackage{subfigure}
\usepackage{psfrag}
\usepackage{url}
\usepackage{color}
\usepackage{multirow}
\usepackage[english]{babel}
\usepackage{mathrsfs}

\usepackage{amsmath,amssymb}

\def\eprinttmp@#1arXiv:#2 [#3]#4@{
\ifthenelse{\equal{#3}{x}}{\href{http://arxiv.org/abs/#1}{#1}
}{\href{http://arxiv.org/abs/#2}{arXiv:#2} [#3]}}

\providecommand{\eprint}[1]{\eprinttmp@#1arXiv: [x]@}
\newcommand{\adsurl}[1]{\href{#1}{ADS}}

\def\bea{\begin{eqnarray}}
\def\eea{\end{eqnarray}}

\begin{document}
\title[Disentangling dark energy and cosmic tests of gravity from weak lensing systematics]
{Disentangling dark energy and cosmic tests of gravity from weak lensing systematics}
\author[I. Laszlo, R. Bean, D. Kirk, S. Bridle] { Istvan Laszlo$^{1}$, Rachel Bean$^{1}$, Donnacha Kirk$^{2}$, Sarah Bridle$^{2}$
\\
$^{1}$Department of Astronomy, Cornell University, Ithaca, NY 14853, USA\\
$^{2}$Department of Physics \& Astronomy, University College London, Gower Street, London, WC1E 6BT, UK}

\pagerange{\pageref{firstpage}--\pageref{lastpage}} \pubyear{2009}

\maketitle

\label{firstpage}

\begin{abstract}
We consider the impact of key astrophysical and  measurement
systematics on constraints on dark energy and modifications to gravity on cosmic scales.
We focus on  upcoming photometric ``Stage III" and ``Stage IV" large scale structure surveys  such as DES, SuMIRe, Euclid, LSST and WFIRST. We illustrate the different redshift dependencies of gravity modifications compared to intrinsic alignments, the main astrophysical systematic.  The way in which systematic uncertainties, such as galaxy bias and intrinsic alignments, are modelled can change dark energy equation of state and modified gravity figures of merit by a factor of four. The inclusion of cross-correlations of cosmic shear and galaxy position measurements helps reduce the loss of constraining power from the lensing shear surveys. When forecasts for Planck CMB  and Stage IV surveys are combined,
constraints on the dark energy equation of state and modified gravity model are recovered, relative to those from shear data with no systematic uncertainties, if fewer than 36 free parameters in total are used to describe the galaxy bias and intrinsic alignment models as a function of scale and redshift. To facilitate future investigations, we also provide a fitting function for the matter power spectrum arising from the phenomenological modified gravity model we consider.
\end{abstract}


\section{Introduction }

Upcoming large scale structure surveys promise to provide deep and wide angular scale measurements of both the distribution of luminous matter, through galaxy position surveys, and the total matter distribution, through weak lensing shear measurement. These include the  ground-based KIlo-Degree Survey (KIDS), Pan-STARRS~\footnote{\tt{http://pan-starrs.ifa.hawaii.edu}}, Subaru Measurement of Images and Redshifts (SuMIRe)~\footnote{\tt{http://sumire.ipmu.jp/en/}} the Dark Energy Survey (DES)~\footnote{\tt{http://www.darkenergysurvey.org}} and the Large Synoptic Survey Telescope (LSST)~\footnote{\tt{http://www.lsst.org}}, and prospective space-based Euclid\footnote{\tt{http://sci.esa.int/euclid}} and WFIRST\footnote{\tt{http://wfirst.gsfc.nasa.gov}} surveys. The expectation is that these will support a significant improvement in constraints on the effect of dark energy  on the growth of cosmic structure.

There has been an active discussion of the implications of such measurements for constraining cosmological models that might predict growth histories that are distinct from those of the `vanilla' $\Lambda$CDM cosmological model, in which the universe is populated by normal matter, cold dark matter (CDM) and a cosmological constant ($\Lambda$) and an evolution determined by General Relativity (GR). Particular attention has been given to models arising from large scale modifications of gravity, such as those proposed as alternatives to explain  the accelerating expansion of the universe, for example \cite{Dvali:2000hr,Sawicki:2005cc,  Zhang:2005vt,Amarzguioui:2005zq,Bean:2006up,Guo:2006ce,Movahed:2007ie, Fang:2008kc,Song:2007da,Carvalho:2008am,Schmidt:2009am,Xia:2009gb,Lombriser:2009xg, Copeland:2006wr,Jain:2010ka,Capozziello:2011yr,Harko:2011kv}.

Rather than considering a specific theory, one can consider a phenomenological parameterisation for how the gravitational metric perturbations are related to the underlying matter distribution and motion, such as was discussed in \cite{Ishak:2005zs,Knox:2006fh,Stab:2006yuk,Kunz:2006ca,Song:2006sa,Laszlo:2007td,Amendola:2007rr,Zhang:2007nk,Linder:2007nu,Jain:2007yk,Bertschinger:2008zb,Zhao:2008bn,Song:2008vm,Thomas:2008tp,Song:2008xd,Zhao:2008bn,Silvestri:2009hh,Caldwell:2009ix,Stril:2009ey,Zhao:2009fn,Guzik:2009cm,Reyes:2010tr,Pogosian:2010tj,Jennings:2010ne,Martinelli:2010wn,Hwang:2010ht,Song:2010fg,Daniel:2010ky,Bean:2010zq,Zhao:2010dz,Daniel:2010yt,Dossett:2011zp,Camera:2011ms,Jain:2011fv,Clifton:2011jh,Linder:2011hi,Tsujikawa:2010zza,Beynon:2009yd}.  One parameterisation in particular
is well suited to contrasting information about galaxy positions with weak lensing measurements.
It
specifies how the evolution of the two scalar metric perturbations in the conformal Newtonian gauge, the ``Newtonian potentials",  is distinct from that predicted by GR.  While galaxy distributions are sensitive to just one of the potentials, lensing is sensitive to the sum of both.

Cross-correlations between position and lensing observables have been proposed as a source of additional cosmological information by \cite{Hu:2003pt} and were used in \cite{Bernstein:2008aq}.
It has been found that these observables could greatly increase constraining power for modified gravity (MG) theories  \citep{Zhang:2007nk,Zhang:2008ba}, however such cross-correlations also play an important role in mitigating critical sources of systematic error, in particular the effect of intrinsic alignments (IA) on the observed shear signal and galaxy bias \citep{Joachimi:2009ez,Kirk:2010zk,Zhang:2010pxa}.

However, since both an alternative theory of gravity and intrinsic alignments qualitatively alter the measured shear signal and its correlations with other observables, one might expect this to translate into a degeneracy between MG and IA parameters, and a deterioration of cosmological constraints. For example, \cite{Zhang:2007nk} highlight the use of galaxy-shear correlations (galaxy-galaxy lensing) to constrain modified gravity models in the absence of intrinsic alignments, whereas \cite{Zhang:2010pxa} proposes the use of the same correlations as a method for measuring intrinsic alignments in GR models.

 In this paper, we characterise and quantify the  degeneracies between dark energy parameter measurements and weak lensing systematics, and assess how they affect our ability to obtain bounds on the nature of cosmic acceleration and gravity on cosmic scales. In a related paper \citep{MGPaper2}, we discuss how surveys can be optimised in light of intrinsic alignment and galaxy bias systematics.

Intrinsic alignments of galaxies arise from the fact that, as large scale structure is assembled, galaxies acquire an intrinsic correlated ellipticity. Cosmic shear is the apparent coherent alignment of galaxies due to gravitational distortions by intervening matter. Therefore, if intrinsic alignments are not accounted for then they can be
falsely attributed to weak lensing and thus contaminate observations. Such intrinsic correlations arise in two forms: an effect from physically close galaxies forming or evolving in the same potential (termed II) and a correlation between the apparent shape of a distant galaxy lensed by a foreground mass and the intrinsic alignment of a foreground galaxy due to the same foreground mass (termed GI) \citep{Hirata:2004gc}.

Including IA effects in a cosmological analysis is well motivated; comparing simulations to observations reveals a need for these effects \citep{Croft:2000gz,Crittenden:2000wi,Heavens:2000ad,Heymans:2006nu,Mandelbaum:2006,Hirata:2007,Okumura:2008bm,Okumura:2008,Faltenbacher:2008fg,Pereira:2010pe} and alignments have been directly observed in existing large scale structure surveys \citep{Brown:2000gt,Hirata:2004jn,Pereira:2004tn,Mandelbaum:2005wv,Agustsson:2005xj,Hirata:2007,Mandelbaum:2009ck,Brainerd:2009da,Siverd:2009vi,Faltenbacher:2008fg,Lee:2010sd,Joachimi:2010xb,Lee:2010sd,Blazek:2011xq,Hao:2011nz}. At typical redshifts, the intrinsic alignment signal can be about ten per cent of the cosmic shear signal we want to measure. However, to measure the dark energy equation of state to percent level accuracy, we need to measure the cosmic shear signal itself to a similar accuracy, therefore incorrectly ignoring the intrinsic alignment signal can lead to biases of order tens of percent in the dark energy equation of state \citep{BridleKing:2007}.  The difficulty in lensing then, even for a perfect instrument, is to separate out the true lensing signal (termed GG) from these intrinsic alignment terms.

The II term can be removed by avoiding inclusion of   physically close  galaxy pairs as in \cite{King:2002,King:2003,Heymans:2003,Takada:2004}. However, no similar simple treatment can be used for GI, since GI depends on line of sight interactions between background alignments and forground lensing, and it is precisely the integrated lensing distortion along the line of sight that is relevant to the pure weak lensing signal. This leaves a  contamination that must be handled in order for weak lensing to improve in quality even as instrumental errors are being beaten down \citep{Hirata:2007}. Recent work has  shown that one can remove some IA contamination by ``nulling'' \citep{Joachimi:2010va,Joachimi:2010sn}. This technique involves a transform of the shear data which downweights the impact of both II and GI terms. Nulling essentially removes information from the survey in order to produce unbiased, but necessarily less tight, constraints. An alternative approach is to model the IA and include additional parameters to characterise the uncertainty in the model.

There has been  some  effort centered on producing a robust theoretical model for IA formation and evolution, primarily derived from the Linear Alignment (LA) model \citep{Catelan:2001,Hui:2002,Hirata:2004gc,BridleKing:2007,Lee:2008}.
These models are based on the assumption that elliptical galaxies take shapes aligned with those of their parent dark matter haloes, which themselves respond linearly to the background gravitational tidal field. Spiral galaxies have shapes produced by torques acting on the angular momentum axis of the galaxies and are shown not to produce IAs to first order.  The LA model is a simple model which generates  results consistent with GI measurements \citep{Joachimi:2011,Blazek:2011}.

Although the linear alignment model serves as a suitable starting point for generating agreement with observations, there are several complicating factors. IAs are likely to be modified through merger and accretion events as structure forms, and also impacted by additional factors such as color, redshift, galaxy type, and luminosity dependence and one-halo corrections \citep{Hirata:2007,Faltenbacher:2008fg,Mandelbaum:2009ck,Hao:2011nz,Lee:2010sd,Schneider:2010,Kirk:2010zk,Joachimi:2011}.

Rather than attempt to model each of these complex effects directly, we allow for added freedom in the LA model by introducing a gridded bias scheme, gridded in scale and redshift space, and marginalise over the additional parameters.  Including such a scheme  to account for uncertainties in the IA amplitude, variation with galaxy type and evolution effects,  will  impact constraints on cosmological parameters  \cite{Bernstein:2008aq,BridleKing:2007,Kitching:2010tp,Joachimi:2011}.

A primary goal of future surveys is to beat down instrumental systematic errors to improve cosmological constraining power. In extracting cosmological constraints, however, one has to also disentangle astrophysical systematic effects. Along with the effect of baryons on small scales, our ability to model the growth of non-linearities in large scale structure and the contamination due to intrinsic alignments are, perhaps, the main astrophysical systematic challenges facing weak lensing surveys. The formation, evolution as a function of redshift and scale, and variation by galaxy type of IAs are all relativity unknown. Realistically one might expect that one will have to extract an understanding of dark energy and IAs simultaneously from lensing and galaxy position correlations, this is particularly so in the context of modified gravity, which affects both the cosmic homogenous expansion and inhomogeneous perturbation growth histories. Here we present a detailed analysis of this task at the level of a Fisher matrix analysis.

The paper is structured as follows: in section \ref{formalism} we outline the formalism used to describe the cosmological model, including dark energy parameters to describe modifications to $\Lambda$CDM in how the homogeneous expansion and growth of inhomogeneities evolve. We describe how the large scale structure and CMB surveys are modeled and how we treat uncertainties in galaxy bias and intrinsic alignments. In section \ref{analysis} we present the results of the Fisher matrix investigation and summarise our findings in \ref{conclusions}. In an Appendix we  introduce a fitting function that allows others to generate weak lensing and galaxy position correlations for the modified gravity model we consider here.

\section{Formalism}
\label{formalism}

In this section we describe the details of our analytical approach: the cosmological model we use to parameterise dark energy and modifications to gravity is outlined in section \ref{cosmomod}, sections \ref{Obs} and \ref{survey} respectively describe the statistical correlations for large scale structure and CMB observables  and  the survey specifications we assume. In \ref{IA} and \ref{grid} we describe how systematic uncertainties in the galaxy bias model and intrinsic alignments  are included in the analysis.
\subsection{Cosmological model}
\label{cosmomod}

Theories that suggest a cosmic scale modification to gravity can produce deviations from $\Lambda$CDM in both the homogeneous expansion history and the growth of inhomogeneities. We model alterations from $\Lambda$CDM in the expansion history using an effective equation of state for the additional physics producing the cosmic acceleration
\bea
w=w_0+w_a(1-a),
\eea
where $a$ is the expansion factor, with $a=1$ today, and $w_0$ and $w_a$ describe the effective  equation of state today and its derivative with respect to $a$ respectively ($w_0=-1$ and $w_a=0$ recover a cosmological constant, $\Lambda$.).

The Friedmann equation relates the Hubble expansion rate, $H(a)$, to the cosmic energy density, $\rho(a)$
\bea
H^2(a) &=& \frac{8\pi G}{3c^2}\sum_i \rho_i(a)
\\
&=& H_0^2\left[\frac{\Omega_m}{a^3}+\frac{\Omega_\gamma}{a^4}+\Omega_\Lambda  a^{-3(1+w_0+w_a)}e^{3w_a(1-a)}\right] \hspace{0.5cm}
\eea
where  $\rho_i$ is the homogeneous (background) density of component $i$, $H_0$ is Hubble's constant, $\Omega_m$  and $\Omega_\gamma $ are the fractional energy densities today in non-relativistic and relativistic matter respectively.

To describe the growth of inhomogeneities we use the conformal Newtonian gauge using the notation of \cite{Ma:1995ey}
\bea
ds^2 = -a^2(\tau)[1+2\psi(x,\tau)] d\tau^2+ a^2(\tau)[1-2\phi(x,\tau)] dx^2
\eea
where $\tau$ is conformal time, $x_i$ are comoving coordinates and $\psi$ and $\phi$ are the Newtonian potentials.

We consider a phenomenological parameterisation for cosmic scale deviations from GR that employs two functions to modify the perturbed Einstein equations. Such modifications have been widely discussed in the literature, both because they are phenomenologically simple, and
also because they can be mapped onto predictions for modifications derived from scalar tensor and higher dimensional theories of gravity.

Unlike with the equation of state, notation for these modifications varies widely. Here we write the modified  Einstein equations as
\bea
k^2\phi &=&  -4\pi G Q a^2\sum_i\rho_i\Delta_i\label{EE000i}
\\
\psi -R\phi&=&  -   12\pi G Q a^2 \sum_i \rho_i(1+w)\frac{\sigma_i}{k^2} \label{EEij}.
\eea
where the first, the Poisson equation, comes from a combination of the time-time and time-space Einstein equations, and the second is the anisotropic space-space equation. Here $k$ is the comoving wavenumber, $\Delta_i$ is the perturbation of component $i$ in the component's rest-frame (the frame in which it has zero peculiar velocity) and $\sigma_i$ is its anisotropic stress.  $\Delta_i$ is a gauge invariant perturbation, related to a perturbation in a general frame, $\delta_i$, in which the component's peculiar velocity is, $v_i$, by $\Delta_i = \delta_i+3\mathcal{H} (1+w_i)v_i/k$. In the discussion below, we focus on density perturbations in CDM, denoted by $\Delta_c$.

In General Relativity the two functions $Q=R=1$. The function $Q(k,a)$ describes the relation between space-space gravitational potential perturbation and a matter overdensity through the Poisson equation, creating an effective gravitational constant $G_{eff}(k,a)=Q(k,a)G$. The second function $R(k,a)$ modifies the relationship between the two Newtonian potentials. In GR, an inequality between these can only be generated by the presence of relativistic matter, through the presence of anisotropic shear stress. Shear stresses are rapidly suppressed in non-relativistic  matter, and would be extraordinarily difficult to sustain in a fluid with a negative equation of state. As such, $R\neq 1$ could be viewed as a potential smoking gun signal of a modification to GR being present. We assume that the fluid equations for normal matter are unaltered and assume that dark energy, as a modification of gravity rather than a fluid, does not cluster.

The growth of overdensities on large and small scales have different dependencies on $Q$ and $R$ \cite{Bean:2010zq}. On small scales, they are purely determined by the peculiar motion during infall, $\psi$, proportional to $QR$,
\bea
\label{eq:smallscale}
\ddot{\Delta}_c+\mathcal{H}\dot{\Delta}_c-\frac{3\mathcal{H}^2}{2}\Omega_m(a) Q R\Delta_c\approx 0,
\eea
where ${\mathcal H}=d\ln a/d\tau=aH$.  For late-time, large scale behaviour the degeneracy between $Q$ and $R$ is not present and CDM density perturbation evolution is governed by
\bea
\label{eq:largescale}
\ddot{\Delta}_c+\mathcal{H}\dot{\Delta}_c-3(\ddot{\phi}+\mathcal{H}(2\dot{\phi}+\dot{\psi})+(2\dot{\mathcal{H}}+\mathcal{H}^2)\psi)\approx 0.
\eea

We assume that time evolution of the effects of modifications to GR on the growth of inhomogeneities would vary in concert with alterations to the background expansion, and consider modifications of the form,
\bea
Q(a) &=& 1+(Q_0-1)a^s
\\
R(a) &=& 1+(R_0-1)a^s.
\eea
For our analysis we fix $s=3$ to allow modifications to the growth of structure to evolve at a comparable rate to the onset of an accelerative component in the homogeneous expansion history. We omit any scale dependence in our modifications. Two conditions, $Q(a)>0$ and $R(a)>-1$ are imposed to ensure that overdensities remain gravitationally attractive and that light is bent towards the lens.

\subsection{Observables}
\label{Obs}

We are interested in how well measurements of weak lensing shear distortions and galaxy counts, from future surveys, will be able to constrain deviations from $\Lambda$CDM. Combining these observables is key to detecting changes in the growth history of the universe as they depend differently on the Newtonian potentials.

We characterise each observable by their 2D angular power spectrum, $C_{\ell}$, for auto- and cross-correlations between each observable
in a given redshift bin.
For two fields $X$ and $Y$, $C_{\ell}$, under the Limber approximation, is given by
\bea
\label{eq:cl}
C^{XY}_{\ell} &= &\int_0^{\chi_{\infty}} \frac{d\chi}{\chi^2}W_X(\chi)W_Y(\chi)S_X(k_{\ell}, \chi)S_Y(k_{\ell}, \chi)
\eea
where $k_{\ell} = \ell/\chi$, and $X,Y$=\{$\delta$, $G$\} for mass and lensing shear fields respectively. $W_X$ and $S_X$ are the window function and source function associated with the field $X$, respectively.

The source functions for the mass distribution and weak lensing shear
are
\bea
S_{\delta} &=&\Delta_c,
\\
S_{G} &=& -\frac{k^2}{2}(\phi+\psi).
\eea
where we have assumed that the density perturbation for matter is equivalent to that for CDM

The power spectrum  for the correlation between fields $X$ and $Y$ is related to the source functions by,  $P_{XY}\equiv \langle S_XS_Y\rangle $, allowing one to rewrite the angular correlation in a common form:
\bea
\label{eq:clpk}
C^{XY}_{\ell} &= &\int_0^{\chi_{\infty}} \frac{d\chi}{\chi^2}W_X(\chi)W_Y(\chi)P_{XY}(k_{\ell}, \chi).
\eea

The galaxy and lensing window functions are dependent on the normalised distribution of galaxy number density in each redshift bin $i$ for the relevant survey, ${n}_{i}(\chi)$. We assume the galaxies are distributed according to \citep{Smail:1994sx} with
\bea
n(z)
\propto z^2 \exp\left(-\frac{z}{z_0}\right)^{3/2}.
\eea
We break up the galaxy distribution into $N_{ph}$ photometric redshift bins, divided so that they each contain an equal number of galaxies.  The photometric redshifts are measured with accuracy $\sigma(z) =\sigma_{z0}(1+z)$ and could have a potential systematic offset, which we model as a constant within each redshift bin, $\Delta z_{i}$. The observed distribution of galaxies in bin $i$ is given by
\bea
n_i(z)
&=& \frac{n(z) }{2}\left[\mathrm{erf}(z_i-z+\Delta z_{i})-\mathrm{erf}(z_{i-1}-z+\Delta z_{i-1})\right]\hspace{0.75cm}
\label{eq:niz}
\eea
The galaxy and lensing window functions are then given by
\bea
W^i_{m}(\chi) &=& \hat{n}_i(z)
\\
W^i_{G}(\chi) &=& \int_{\chi}^{\chi_{\infty}} d\chi' \hat{n}_i(\chi') \frac{r(\chi)r(\chi'-\chi)}{r(\chi')},
\eea
where $r(\chi)$ is the comoving angular diameter distance to comoving distance $\chi$, and $\hat{n}_i$ is the normalised number density,
\bea
 \hat{n}_i(z)=\frac{n_i(z)}{\int_{z=0}^{\infty}n_i(z)dz}.
\eea

In relating  the observed correlations of galaxies to the dark matter correlation functions above, we must account for a bias between dark and luminous matter. This bias is dependent on the galaxy type, redshift and environment in a way that is poorly understood. We allow for this uncertainty by introducing a redshift and scale dependent bias parameter, $b_g$, to relate the auto-correlation of the galaxies to the autocorrelation of the mass and an independent, correlation parameter, $r_g$, to describe the bias in the cross-correlation of luminous matter and the mass.
The correlation parameter, $r_g$, would be equal to unity if the galaxies trace the mass, and less than unity if there is some stochasticity in the galaxy formation process e.g. see \cite{Dekel:1998eq} for a review.

The galaxy position auto-correlation and galaxy position-shear cross-correlation are then related to the underlying mass and shear observables by
\bea
P_{gg}(k,\chi)&=& b_g^2(k,\chi)P_{\delta\delta}(k,\chi),\\
P_{gG}(k,\chi)&=& b_g(k,\chi)r_g(k,\chi)P_{\delta G}(k,\chi).
\eea
where $g$  and $\delta$, denote galaxy and underlying mass observables respectively.

Note that in modified gravity models the lensing source term and the mass source term must be
different
\bea
P_{\delta G}(k,\chi)&=& \left[\frac{Q(\chi)(R(\chi)+1)}{2}\right]P_{\delta\delta}(k,\chi),
\\
P_{GG}(k,\chi) &=& \left[\frac{Q(\chi)(R(\chi)+1)}{2}\right]^2P_{\delta\delta}(k,\chi). \label{eq:PGG}
\eea
The growth of the dimensionless power spectrum $P_{\delta\delta}$ is itself dependent on modified gravity parameters $Q$ and $R$, as summarised by (\ref{eq:smallscale}) and  (\ref{eq:largescale}).

To obtain the lensing and galaxy position correlations in the modified gravity scenarios we integrate the full equations of motion using a modified version of CAMB \citep{Lewis:1999bs}.

To support other researchers investigating the role of modified gravity models on large scale structure observations, without having to integrate the full perturbation equations, we provide a fitting function  in  the Appendix
for the ratio, $r_{fit}(k,z)$, between a fiducial $\Lambda$CDM linear matter power spectrum,
$P_{\delta\delta,\Lambda CDM}(k,z)$ and the one for a modified gravity model described in \ref{cosmomod}, parameterised by $Q_0,R_0$ and $s$:
\bea
r_{fit}(k,z;Q_0,R_0,s) \equiv\frac{P_{\delta\delta,fit}(k,z;Q_0,R_0,s) }{P_{\delta\delta,\Lambda CDM}(k,z)}.
\eea

\subsection{Survey specifications}
\label{survey}
We consider the impact of including IAs on cosmological constraints for a near-term Dark Energy Task Force (DETF) \cite{DETF} Stage III  survey, such as DES
or SuMIRe, and a longer-term Stage IV survey, such as Euclid, LSST or WFIRST.
\begin{table}
\begin{tabular}{|c|c|c|}
	\hline
Survey Parameters&Stage III & Stage IV\\
\hline
{Area(sq. deg.)} &5000& 20000\\

{$\sqrt{2} z_0$} &0.8& 0.9 \\

{$z_{min}$} & 0.001 &0.001\\

{$z_{max}$} & 3 &3\\

$N_g$ & 10& 35 \\

$N_{ph}$ & 5& 10 \\

$\sigma_{z0}$  &0.07&0.05\\
$\gamma_{rms}$ &$0.23$& $0.35$ \\
\hline
\end{tabular}
\caption{Summary of the photometric large scale structure survey specifications assumed for the Stage III and Stage IV survey: survey area; median survey redshift, $\sqrt{2} z_0$; minimum and maximum redshifts observed, $z_{min}$ and $z_{max}$; number of galaxies, per square arcminute, $N_g$; number of photometric redshift bins, $N_{ph}$; standard photometric redshift measurement error at $z=0$, $\sigma_{z0}$, and the r.m.s. shear measurement error, $\gamma_{rms}$. }
\label{table-LSSpar}
\end{table}

\begin{table}
\begin{tabular}{|c |c| c| c|}

\hline
$\nu$(GHz) & 100 & 143 & 217\\
\hline
\multicolumn{1}{|c|}{$f_{sky}$} &0.8 & 0.8 & 0.8\\
\multicolumn{1}{|c|}{$\theta_{FWHM}$(arc min)} & 10.7 & 8.0 & 5.5\\
\multicolumn{1}{|c|}{$\sigma_{T}$($\mu$K)} & 5.4 & 6.0 & 13.1\\
\multicolumn{1}{|c|}{$\sigma_{E}$($\mu$K)} & - & 11.4 & 26.7\\
\hline
\end{tabular}
\caption{CMB survey specifications for a Planck-like survey. We model this on the temperature, $T$, and $E$-mode polarisation specifications from three lowest frequency bands for the Planck HFI instrument. }
\label{table-Planckpar}
\end{table}


The noise for each survey is modeled as statistical errors given by
\bea
N_{\ell}^{\epsilon_i\epsilon_j}&=&\delta_{ij}\frac{\gamma_{rms}^2}{2n_j},
\\
N_{\ell}^{n_in_j}&=&\delta_{ij}\frac{1}{n_j},
\\
N_{\ell}^{n_i\epsilon_j}&=&0,
\eea
where $\gamma_{rms}$ is the root mean square uncertainty in the shear measurement of the galaxies and $n_j$ is number of galaxies per steradian in $j^{th}$ photometric redshift bin so $\sum_i n_i = N_g$.

The survey specifications assumed in our analysis for the Stage III and IV surveys are given in Table \ref{table-LSSpar}.

We include complementary constraints from temperature (T) and E-mode polarisation (E) measurements from a Planck-like CMB survey up to $l=3000$. As summarised in Table \ref{table-Planckpar}, we model this by considering the three lowest frequency bands of the Planck HFI instrument, three channels for temperature data and 2 for E mode polarisation,as described in the Planck Bluebook \footnote{$www.rssd.esa.int/SA/PLANCK/docs/Bluebook-ESA-SCI(2005)1\_V2.pdf$}. We assume each frequency channel has Gaussian beams of width $\theta_{FWHM}$ and error in $X=T,E$ of $\sigma_X$, so that the noise in channel $c$ is given by
\bea
N_ \ell ^{XX,c} &=& \left(\sigma_{X,c} \theta_{FWHM,c}\right)^2 e^{\ell(\ell +1)\theta^2_{FWHM,c}/8\ln(2)},
\eea
and over all channels,
\bea
N_ \ell ^{XX} &=& \left[\sum_{c}\left(N_{\ell,c}^{XX}\right)^{-1}\right]^{-1}.
\eea

\subsection{Intrinsic Alignments}
\label{IA}

Cosmic shear describes the distortion of the image of a distant galaxy due to the bending of light from that galaxy by gravity as it passes massive large-scale structure. For a galaxy in the $i^{th}$ photo-z bin, the observed ellipticity, $\epsilon$, of the galaxy can be written as a sum of three independent contributions:  the cosmic shear $\gamma_G$, the intrinsic, non-lensed shape of the galaxy, $\gamma_I$, and apparent ellipticity introduced through instrumental and foreground noise, $\epsilon_{rnd}$,
\bea
\epsilon^{i}(\theta) &=& \gamma_{G}^{i}(\theta)+\gamma_{I}^{i}(\theta)+\epsilon^{i}_{rnd}(\theta).
\eea

The cosmic shear signal $\gamma_{G}$ is very small, and we cannot measure directly the intrinsic shear of any individual galaxy. To recover the cosmic shear, therefore, one averages over a number of galaxies on a small patch on a sky. Assuming that their intrinsic ellipticities are distributed randomly, and that their light passes by similar large scale structure, the intrinsic ellipticities cancel in the two-point function, and we are left with the cosmic shear signal.

In reality, the assumption that intrinsic ellipticities are randomly distributed on the sky is inaccurate. There are two strains of intrinsic alignment of galaxy ellipticities, both arising from the same physics of galaxy formation.

The measured weak lensing signal reflects a correlation in shapes arising from distant galaxies passing near the same foreground gravitational lens. However, if the background images are already correlated, this boosts the measured signal and gives rise to a systematic deviation in the observed shear.

In section \ref{LA}, we describe the linear alignment model \citep{Catelan:2001,Hirata:2004gc} we use as our basis to describe intrinsic alignment contributions. In this model, galaxies quickly align
along the curvature of the gravitational potential. While this model does provide a reasonable starting point for an overall agreement with observations, and thus a suitable baseline for use in our statistical analysis, it neglects known complicating factors of dependency on galaxy luminosity, galaxy type and redshift, and the effect of post-processing of IA in mergers and accretion events \citep{Hirata:2007,Faltenbacher:2008fg,Mandelbaum:2009ck,Hao:2011nz,Schneider:2010,Joachimi:2011,Lee:2010sd,Blazek:2011}. In section \ref{IAext} we therefore extend beyond the LA model, in an attempt to allow for these effects, through the inclusion of additional non-linear corrections, and scale and redshift dependent bias terms.

\subsubsection{The Linear Alignment Model}
\label{LA}

The Linear Alignment (LA) model introduced in \cite{Catelan:2001,Hirata:2004gc}  assumes that galaxies would align with the stretching axis of the potential in which they form, so that the intrinsic shear
is assumed to be
\bea
\gamma^{(1,2)}_{I}
=-\frac{C_f}{4 \pi G}(\nabla^2_x-\nabla^2_y,\nabla_x\nabla_y)\psi(z_f).
\eea
$\psi(z_f)$ is the smoothed gravitational potential field sourcing the shear alignments at a primordial redshift, $z_f$,when galaxy formation occured and $C_f$ is a normalisation determined by matching the model to observations. We take $z_f=50$ here, however we expect $z_f$ to be well within the matter dominated era, when the potential $\psi$ remains roughly constant in time and gravity can be well-described by GR, so that the analysis should be largely insensitive to the precise value of $z_f$ assumed.

The presence of intrinsic alignments alters the observed shear correlations.  We denote the {\it measured} galaxy shear and position observables by $\epsilon$ and $n$, to distinguish them from the underlying theoretical IA-free shear and position variables, $G$ and $g$ respectively. The observed angular correlation functions for galaxy shear and position, when intrinsic alignments are included, are given by
\bea
C_{\ell}^{\epsilon\epsilon}&=&C_{\ell}^{GG}+C_{\ell}^{GI}+C_{\ell}^{II},\\
C_{\ell}^{n \epsilon}&=&C_{\ell}^{gG}+C_{\ell}^{gI}
\eea
ignoring magnification terms \citep{Joachimi:2009ez}.

The Intrinsic-Intrinsic ($II$) alignment correlation applies to physically close galaxies. These form in the same large-scale gravitational potential and their intrinsic ellipticities tend to align with the field lines of that potential. When these galaxies are observed on the sky they will tend to point in the same direction. This alignment produces a spurious correlation which adds to the observed cosmic shear signal.

Since the $II$ correlation is greatest for closely positioned galaxies, there is a related correlation between the position and the intrinsic alignment of a pair of physically close galaxies, this is the galaxy position-intrinsic alignment ($gI$) correlation.

Somewhat more subtle is the Gravitational-Intrinsic alignment ($GI$) correlation which applies to galaxies close on the sky but separated in redshift. The intrinsic ellipticity of the foreground galaxy will tend to align with the nearby gravitational potential which, in turn, is responsible for gravitationally lensing the background galaxy. This tends to produce an anti-correlation which subtracts from the observed cosmic shear signal as we observe the galaxies oriented orthogonally.

We can write angular correlation functions involving intrinsic alignments using the formalism in (\ref{eq:cl}), defining a source term, $S_I$, and window function, $W_I$, for the intrinsic alignments
\bea
S_{I}(k,\chi)&=& -\frac{C_f }{4 \pi G}k^2{\psi}(k,z_f),\label{SIdef}
\\
W_I^i(\chi) &=& W_m^i(\chi) = \hat{n}_i(\chi).
\eea
The angular correlation is often written in terms of the linear power spectra,  $P_{XY}=\langle S_X S_Y\rangle$,
\bea
C_{ij}^{GI}(l)&=&\int_0^{\chi_{\infty}} \frac{d\chi}{\chi^2}W_{G,i}(\chi)\hat{n}_j(\chi) P_{GI}(k,\chi), \nonumber
\\
C_{ij}^{II}(l)&=&\int_0^{\chi_{\infty}} \frac{d\chi}{\chi^2}\hat{n}_i(\chi)\hat{n}_j(\chi) P_{II}(k,\chi), \label{Cliidef}
\\
C_{ij}^{gI}(l)&=&\int_0^{\chi_{\infty}} \frac{d\chi}{\chi^2}\hat{n}_i(\chi)\hat{n}_j(\chi) P_{gI}(k,\chi).\nonumber
\eea
One can write the correlations in (\ref{Cliidef}) in terms of the matter power spectrum at $z_f$,
\bea
P_{GI}(k,z) &=&\frac{Q(k,z)[1+R(k,z)]}{2}\frac{D(k,z)}{D(k,z_f)}C_f  \bar\rho_{m}(1+z_f)
P_{\delta\delta}(k,z_f), \nonumber
\\
P_{II}(k,z) &=&C_f^2  \bar\rho_{m}^2(1+z_f)^{2}
P_{\delta\delta}(k,z_f), \label{grPII}
\\
P_{gI}(k,z) &=&b_g(k,z)\frac{D(k,z)}{D(k,z_f)}C_f  \bar\rho_{m}(1+z_f)P_{\delta\delta}(k,z_f), \nonumber
\eea
where $\bar\rho_m$ is the mean matter density today and $D(k,z)$ is the linear growth factor for CDM perturbations,
\bea
\Delta_{c}(k,z) =\frac{ D(k,z)}{D(k,z_f)}\Delta_{c}(k,z_f).
\eea

 In GR the linear growth factor is scale independent. If gravity is modified on cosmic scales, however,  it can be scale dependent, and sensitive to the functions $Q$ and $R$.

How does one normalise the IA correlations, and the constant $C_f$? \cite{BridleKing:2007} provide a numerical value for $C_{\ell}^{II}$,  by comparing with \cite{Hirata:2004gc} who used observations of low redshift galaxies today. Rather than normalising  the IA source function at $z_f$, they normalise it, with constant $C_1$, relative to the gravitational potential today assuming a $\Lambda$CDM cosmology, $\psi_{\Lambda}$,
\bea
S_{I}(k,z)&=& -\frac{C_1 }{4 \pi G}k^2\psi_{\Lambda}(k,0).
\eea
 They find  $C_1  = 5\times 10^{-14} (h^2 M_{sun}/Mpc^{-3})^{-1} =  8.25h^{-2}\times 10^{4}  Mpc^2$.
We extrapolate the normalisation of \citep{BridleKing:2007} at $z=0$ to $z=z_f$ by assuming a $\Lambda$CDM growth factor, $D_{\Lambda}(z)$  \citep{Carroll:1991mt}. The early and late time normalisations, $C_f$ from $C_1$, are then related by
\bea
C_f &=&C_1\frac{D_{\Lambda}(0)}{D_{\Lambda}(z_f)(1+z_f)}.\label{norm}
\eea

 \subsubsection{Generalising the IA model}
 \label{IAext}

\begin{figure*}
\center
\includegraphics[scale=0.45]{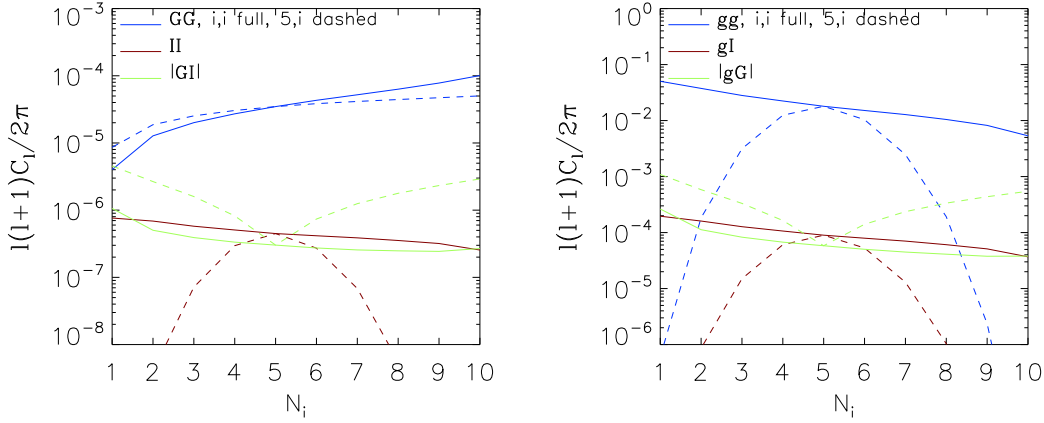}
\caption{
A comparison of the intrinsic alignment and cosmological contributions, assuming a fiducial $\Lambda$CDM cosmology, to the shear-shear [left panel] and position-shear correlations [right panel] as a function of photometric redshift bin, $N_i$, for the Stage IV survey specification for a single multipole, $\ell=1000$. Same-bin `$ii$' [full lines] and cross-bin correlations with the 5$^{th}$, central, redshift bin  `$5i$' [dashed] are shown for the cosmological correlations $GG$, $gg$ and $gG$, and the intrinsic aligment correlations $II$, $GI$, and $gI$. For the $gG$, $gI$ and $GI$ correlations we plot the larger of i5 and 5i correlations in each case. }
\label{fig:spectrabyz}
\end{figure*}

\begin{figure*}
\center
\includegraphics[scale=0.45]{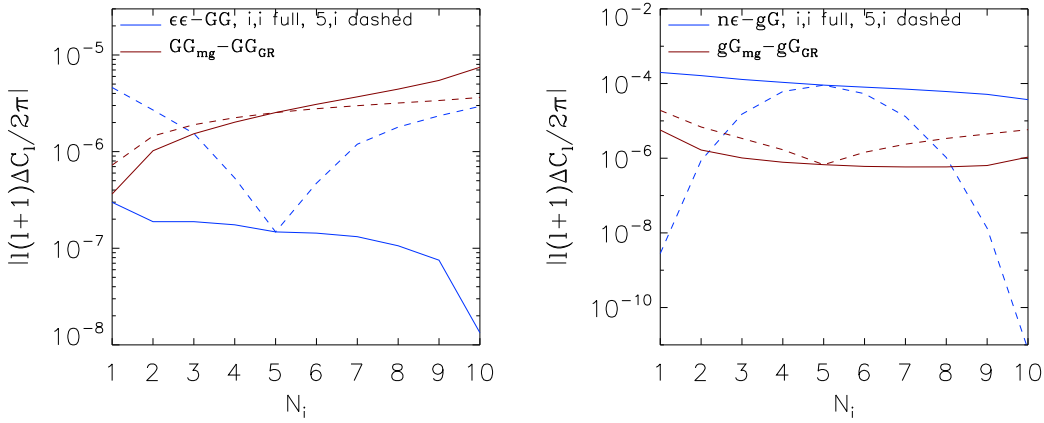}
\caption{
The difference in shear-shear [left panel] and position-shear correlations [right panel] in which IAs are included [blue lines] or using a modified gravity model (with no IAs) [red lines] in comparison to a fiducial model, in which no IAs are included and GR is assumed. The modified gravity model shown has $Q_0=1.05$ and $R_0=1$.  As in figure \ref{fig:spectrabyz}, we show  correlations for  $\ell=1000$ in each photometric redshift bin, $N_i$, for the Stage IV specification.}
\label{fig:spectradiffbyz}
\end{figure*}

Though the correlations in galaxy orientation might be formed in the era of galaxy formation, the intrinsic alignments we observe will invariably be sensitive to the evolution of those galaxies, and the galaxy environment.

One factor that is not included in the LA model is the impact of non-linear clustering of galaxies on the distribution of the galaxies sourcing the intrinsic alignments. A non-linear alignment (NLA) model was introduced as an ad-hoc way to incorporate non-linear clustering into the LA model \citep{Hirata:2007,BridleKing:2007}. This replaces the linear power spectrum in the intrinsic alignment angular correlation expression (\ref{grPII}) with the non-linear power spectrum based on the fitting function derived from the halo model of \cite{Smith:2002dz}.

While this is somewhat adhoc,  it was found to give a qualitatively similar result to the more motivated halo model of intrinsic alignments in \cite{Schneider:2010}
and it has been shown,  in for example  \cite{Hirata:2007,Blazek:2011}, that it gives a more consistent fit to the data than the linear alignment model.

We extend this approach to take into account the modified growth history, if gravity deviates from GR. We model the effect of non-linear clustering by boosting the IA source function, $S_I^{NLA}$ relative to that predicted by linear alignments,
\bea
S_I^{NLA}(k,\chi)=S_I^{LA}(k,\chi) \sqrt{\frac{P_{\delta\delta}(k,\chi)^{nonlin}}{P_{\delta\delta}(k,\chi)^{lin}}}.
\eea
here $P_{\delta\delta}^{lin}$ is the linear matter power spectrum predicted modified gravity model, and $P_{\delta\delta}^{nonlin}$ is the non-linear spectrum  after a correction is applied to the power spectrum using the Smith et. al. halo fitting function \citep{Smith:2002dz}. The use of the Smith et al. fitting function, for a modified expansion history, is equivalent to assuming that non-linear collapse in the modified gravity theories follows the Zel'dovich approximation; this was shown to be reasonable if the phenenomological modifications in (\ref{EE000i}) and (\ref{EEij}) hold to nonlinear scales \citep{Stab:2006yuk,Laszlo:2007td}. We briefly discuss the motivation and possible impact of deviations from this assumption later in the analysis.

In figure \ref{fig:spectrabyz} we show how each intrinsic alignment contribution to the observed correlations varies as a function of redshift for a fixed multipole, $\ell=1000$. We refer the interested reader to \cite{Joachimi:2009ez} for a figure detailing the variation in lensing, galaxy and IA correlations across all multipoles and redshift bins. The same-bin correlations vary monotonically as a function of redshift (denoted by the index of the photometric redshift bin), while the cross-bin correlations involving galaxy positions dramatically fall off as the photometric redshift bins become more separated. By contrast, the cross-bin correlations for $GG$ and $GI$, can remain significant even in cross-correlations between widely separated bins because of the broad redshift kernel for the lensing window function.

Figure \ref{fig:spectradiffbyz}  compares the  amplitude and variation in shear and galaxy correlations when, separately, IAs are added and when the modified gravity theory is allowed. While the two effects can be of comparable amplitude, their distinct redshift dependencies, if the intrinsic alignment model is perfectly known, could assist in disentangling them.

Intrinsic alignments are unlikely, however,  to be described fully by the linear alignment model; they will depend on the details of galaxy formation within dark matter halos, and baryonic physics within galaxies, with complexity beyond this model. For example, it is known that the IA signal depends strongly on galaxy type and color, \citep{Lee:2010sd,Blazek:2011}; spirals are supported by angular momentum, and so more likely subject to tidal torquing of the angular momentum vector, while elliptical galaxies are better described by linear alignments. This bifurcation translates into colour dependence, as spirals are blue while ellipticals are older and redder, and is noted in surveys which split samples by colour \citep{Hirata:2007,Faltenbacher:2008fg,Mandelbaum:2009ck}. Redshift and luminosity dependences, and one-halo corrections at smaller scales also exist, as in \citep{Hirata:2007,Faltenbacher:2008fg,Mandelbaum:2009ck,Hao:2011nz,Schneider:2010,Joachimi:2011}.

Rather than attempt to incorporate these numerous effects by direct modelling, we choose to allow a scale and redshift dependent bias factor to parameterise our ignorance  and marginalise over the bias parameter in a redshift and scale gridding. We introduce two additional bias parameters into the IA correlations, $b_I$ and $r_I$, analogous to the galaxy bias $b_g$ and $r_g$, to reflect our uncertainty in the bias model:
\bea
P_{II}&=& b_I^2P_{II}^{(fid)}
\\
P_{GI}&=& b_Ir_IP_{GI}^{(fid)}
\eea
where the fiducial functions are given by the LA or NLA mode.

 \subsection{Modeling galaxy bias and IA amplitudes}
 \label{grid}

We consider two scenarios for galaxy and IA bias. In our simple model all bias parameters are $k$ and $z$ independent, i.e. $b_g$, $b_I$, $r_g$ and $r_I$ are each a single constant free parameter determining the amplitude of the bias. In the more realistic scenario, motivated by \cite{Joachimi:2009ez} each bias coefficient, $B_X\in\{b_g,r_g,b_I,r_I\}$, is interpolated from a $N_k\times N_z$ grid of values logarithmically spaced in $k$ and $z$, $B^{ij}_{X}$, respectively,
\bea
B_X(k,a)&=& (1-\Delta_k)\left[(1-\Delta_z)B_X^{ij}+\Delta_z B_X^{ij+1}\right]\nonumber
\\
&&+\Delta_k\left[(1-\Delta_z)B_X^{i+1j}+\Delta_z B_X^{i+1j+1}\right] \label{biasgrid}
\eea
for $k_i<k<k_{i+1}$ and $z_j<z<z_{j+1}$, with
\bea
\Delta_k &\equiv& \frac{\ln(k/k_i)}{\ln(k_{i+1}/k_i)} \label{gridk}
\\
\Delta_z & \equiv& \frac{\ln[(1+z)/(1+z_j)]}{\ln[(1+z_{j+1})/(1+z_j)]} \label{gridz}
\eea
and modulated by a free constant amplitude parameter. We choose $k_{min}=10^{-3}Mpc^{-1}$ and $k_{max}=30 Mpc^{-1}$ for the gridding, and assume $B_X(k<k_{min})=1$ and $B_{X}(k>k_{max})=B_{X}^{N_kN_z}$ at all times. We consider scenarios in which $1\le N_k=N_z\le 5$. Each of the $N_k \times N_z$ grid nodes is a freely varying parameter. This means that, in our more sophisticated model, there are $4  N_k N_z $ nuisance parameters when all correlations and cross-correlations are included.

A multi-bin marginalisation over bias parameters is arguably conservative, however we believe it reasonably reflects the current uncertainties in the bias and IA models.
It was inspired by the work of \cite{Bernstein:2008aq} which was used in \cite{Albrecht:2009ct} who bin the biases in redshift and multipole bins, rather than redshift and wavenumber as we do here.

For observables involving the galaxy position correlations we truncate the maximum $\ell$ used as a function of redshift bin, as per \cite{Rassat:2008ja,Joachimi:2009ez} to remove poorly understood biasing on
non-linear scales from the likelihood calculation. We introduce a maximum wavenumber $k_{max}$ for a photometric redshift bin $i$, and neglect all $\ell_i > k_{max}\chi(z_i)$.

Our analysis spans from the optimistic to conservative scenarios. Optimisitically one might assume one can extract, and therefore exclude, IAs with perfect precision and can model the galaxy bias as scale and redshift independent.A conservative perspective would be to represent our ignorance in IA and galaxy bias modeling with 100 marginalised parameters ($N_k=N_z=5$). Where, within this range, the realistic range will finally fall will  depend on progress in  understanding IAs potentially through the use of complementary spectroscopic redshift surveys and the development of galaxy training sets, or preferential selection of galaxy subgroups in which intrinsic alignments are less pronounced, and galaxy bias is well-understood.

\section{Analysis}
\label{analysis}

For our analysis of the impact of systematics on dark energy constraints, we consider constraints on 10 cosmological parameters:
\bea
{\bf p}&=&\{\Omega_\Lambda,w_0,w_a,Q_0,\frac{Q_0(1+R_0)}{2},  \Omega_bh^2,\Omega_mh^2,\tau_{reion},\nonumber
\\
 && n_s,\ln(10^{10} A_s)\}
 \eea
where $\tau_{reion}$ is the optical depth to the epoch of reionisation, and $n_s$ and $A_s$ are the spectral index and normalisation of the primordial spectrum of curvature perturbations, with pivot scale $k=0.05Mpc^{-1}$. We choose fiducial values for these parameters assuming $\Lambda$CDM, and consistent with a $WMAP$7 bestfit cosmology \citep{Larson:2010gs}.

As can be seen in \ref{Obs}, a primary constraint on modified gravity parameters from weak lensing is the combination from $Q_0(1+R_0)/2$, rather than $R_0$. We therefore  use $Q_0$ and $Q_0(1+R_0)/2$ as variables in the Fisher analysis, and take GR with $Q=R=1$ as the fiducial model.

Unless stated otherwise, we consider a conservative scenario for astrophysical systematics, and marginalise over the galaxy and intrinsic alignment biases $b_g,r_g,b_I,r_I$ with $N_k=N_z=5$ bins. We assume $b_g=r_g=b_I=r_I=1$ for the fiducial model.

Our fiducial scenario involves a $5\times 5$ grid of 4 auto-correlation and cross-correlation galaxy and IA bias parameters and 10 cosmological parameters, giving a total of 110 parameters.

The large number of parameters, especially when using the full bias model, favors the use of a Fisher matrix approach. For $N$ parameters, only $N+1$ samples are required to estimate the parameter covariance matrix, $C_{ij}=F_{ij}^{-1}$, with,
\bea
 F_{ij}=\sum_{ab}\sum_\ell \frac{\partial {\mathscr D}_{a}(\ell)}{\partial p_i}Cov_{ab}^{-1}\frac{\partial {\mathscr D}_{b}(\ell)}{\partial p_j},
\eea
where ${\bf {\mathscr D}}(\ell)=\{C_{\ell}^{CMB},C_{\ell}^{n_in_j},C_{\ell}^{n_i\epsilon_j},C_{\ell}^{\epsilon_i\epsilon_j}\}$ are the set of observables across all multipole bins, and redshift bin combinations.
We consider correlations for $10\leq\ell\leq 3000$ in 50 logarithmically spaced bins in $\ell$ space.

To calculate the partial derivatives,$\partial {\mathscr D}(\ell)/\partial p  $, we take a 2\% reduction in each parameter with non-zero fiducial value, and an absolute step of $-0.02$ for all parameters whose fiducial value is zero. We checked that the results are insensitive to the exact size of the step size, obtaining consistent results with 1\% and 3\% step sizes.

The covariance matrix  $Cov_{ab}^{-1}$ between two observables, in multipole bin with mid-value $\ell$ and spanning $\ell_{min}(\ell)\le \ell'\le\ell_{max}(\ell)$, is given by
\bea
Cov[C_{\ell}^{W_iX_j},C_{\ell}^{Y_mZ_n}] &=&\frac{ \hat{C}_{\ell}^{W_iY_m}\hat{C}_{\ell}^{X_jZ_n}+\hat{C}_{\ell}^{W_iZ_n}\hat{C}_{\ell}^{Y_mX_j}}{f(\ell)f_{sky}} \nonumber
\\
\hat{C}_{\ell}^{W_iX_j} &\equiv& C_{\ell}^{W_iX_j} +N_{\ell}^{W_iX_j} \nonumber
\\
f(\ell) & \equiv& \sum_{\ell'=\ell_{min}(\ell)}^{\ell_{max}(\ell)}(2\ell'+1).
\eea
We have modified the publicly available CosmoMC \citep{Lewis:2002ah} and CAMB \cite{Lewis:1999bs} codes to calculate the Fisher matrix and the correlation functions for the future survey specifications, in light of the IA and modified gravity models.


\begin{figure*}
\center
\includegraphics[scale=0.6]{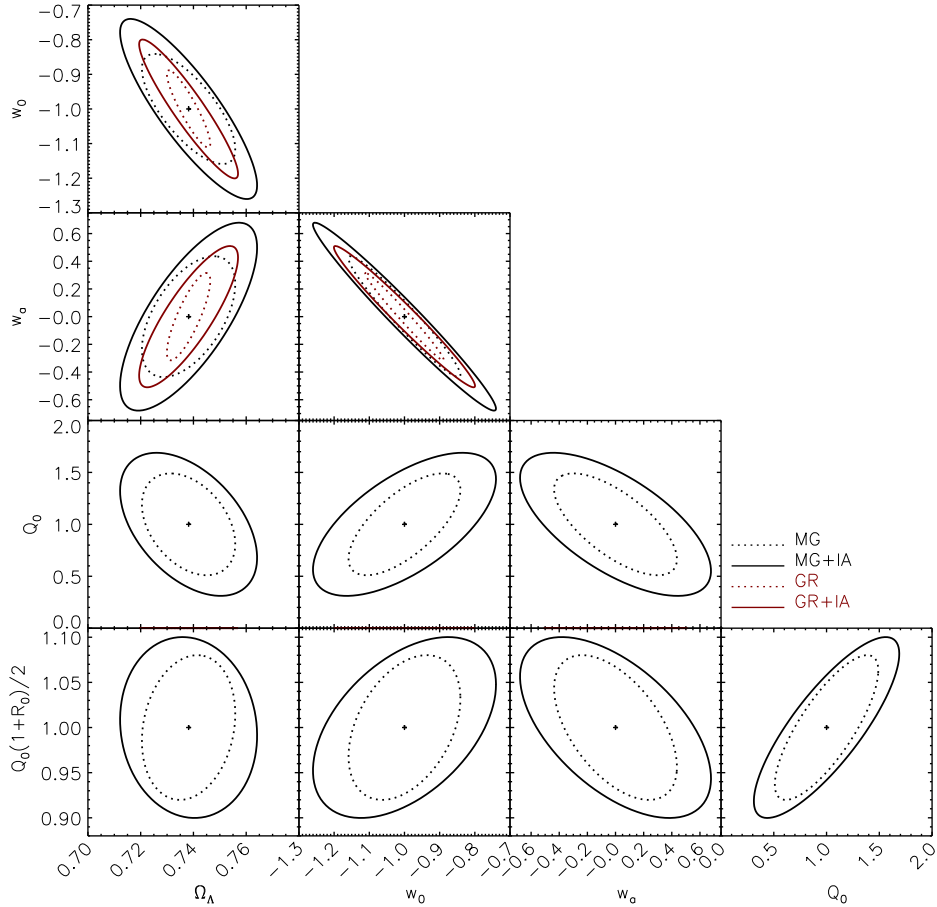}
\caption {A comparison of dark energy constraints in the 2D marginalised parameter planes when intrinsic alignments are included using the LA model [full lines],  in comparison to when it is assumed that IA's are perfectly understood and can be extracted to reveal the underlying cosmological shear and galaxy position correlations [dotted lines]. 95\% confidence level constraints are shown when both GR is assumed [red] and when large scale modifications to gravity ``MG" are allowed [black]. These results combine Planck-like CMB data with a Stage IV survey's galaxy position and shear auto and cross correlations.}
\label{fig:2D}
\end{figure*}

 \begin{table*}
 \label{table-results}
 \begin{center}
 \begin{tabular}{|l|l|c|c|c|c|c|c|c|c|c|c|}
 \hline
 Survey &Scenario       & $\sigma(\Omega_\Lambda)$ & $\sigma(w_0)$ & $\sigma(w_a)$& $\sigma(w_p)$& $FoM(EoS)$ & $\sigma(Q_0)$ & $\sigma\left(\frac{Q_0(1+R_0)}{2}\right)$& $FoM(MG)$& $FoM(comb)$ & $|r_{corr}|$  \\ \hline
 \multicolumn{11}{l}{\it{Assuming GR}}

 \\ \hline
  \multirow{3}{*} {Stage III}                                                & no IA           &     0.014  &     0.169  &     0.451  &     0.029  &      76.1  &      &      & &&   \\
               & LA              &     0.025  &     0.272  &     0.677  &     0.042  &      35.2  &       &       &  && \\
               & NLA            &     0.027  &     0.294  &     0.728  &     0.043  &      32.3  &       &    &   &&\\
 \cline{1-12}
        \multirow{3}{*} {Stage IV}     & no IA           &     0.003  &     0.045  &     0.127  &     0.008  &    1041.1  &      &     & &&   \\
                                                    & LA              &     0.008  &     0.081  &     0.207  &     0.017  &     287.3  &      &     &  && \\
                                                     & NLA            &     0.008  &     0.086  &     0.217  &     0.017  &     268.7  &       &      &  && \\
\hline
 \multicolumn{11}{l}{\it{Allowing an alternate modified gravity model}}
 \\ \hline
 \multirow{3}{*} {Stage III}        & no IA           &     0.019  &     0.226  &     0.605  &     0.034  &      48.8  &     0.666  &     0.091  &      40.1  & 54.9& 0.59\\
                                                 & LA              &     0.030  &     0.363  &     0.923  &     0.043  &      25.3  &     0.894  &     0.118  &      25.0 & 29.6& 0.53\\
                                                   & NLA            &     0.032  &     0.383  &     0.971  &     0.043  &      23.8  &     0.897  &     0.118  &      23.7 & 27.7& 0.51\\
 \cline{1-12}
       \multirow{3}{*} {Stage IV}  & no IA           &     0.007  &     0.064  &     0.176  &     0.017  &     334.5  &     0.198  &     0.032  &     253.6 & 469.8&0.79\\
                                                   & LA              &     0.010  &     0.105  &     0.274  &     0.022  &     166.1  &     0.280  &     0.041  &     151.6& 202.1& 0.62\\
                                                      & NLA            &     0.011  &     0.106  &     0.276  &     0.022  &     164.0  &     0.292  &     0.041  &     143.7& 190.8& 0.60\\
 \cline{1-12}
        \multirow{2}{*} {Stage IV}& no IA           &     0.008  &     0.072  &     0.204  &     0.024  &     200.5  &     0.226  &     0.039  &     196.3 & 327.9&0.78\\
         & LA              &     0.012  &     0.111  &     0.295  &     0.031  &     109.9  &     0.290  &     0.045  &     129.8 & 162.7&0.64\\
         +sys. offsets         & NLA            &     0.012  &     0.112  &     0.296  &     0.031  &     108.4  &     0.301  &     0.046  &     123.8 & 153.9&0.61\\
 \hline
 \end{tabular}
 \caption{
 Comparison of figures of merit and marginalised 1-$\sigma$ errors for dark energy equation of state (EoS) parameters $\{w_0,w_a\}$ and modified gravity (MG) parameters $\{Q_0,Q_0(1+R_0)/2\}$. A combined figure of merit including covariances between all 4 dark energy parameters, $FoM(comb)$, and a correlation coefficient, $r_{corr}$, between the EoS and MG parameters are also included. The table shows prospective constraints from galaxy position and weak lensing auto- and cross-correlations from Stage III and Stage IV surveys in combination with temperature and polarisation data from a Planck-like CMB survey.  We assume a conservative model for galaxy and IA bias parameters, with $N_{k}=N_z=5$. The results with ``+sys. offsets" include marginalisation over weak lensing calibration and photometric redshift offset parameters meant to reflect possible instrumental systematic errors.}
 \end{center}
 \end{table*}

\begin{figure*}
\center
\includegraphics[scale=0.6]{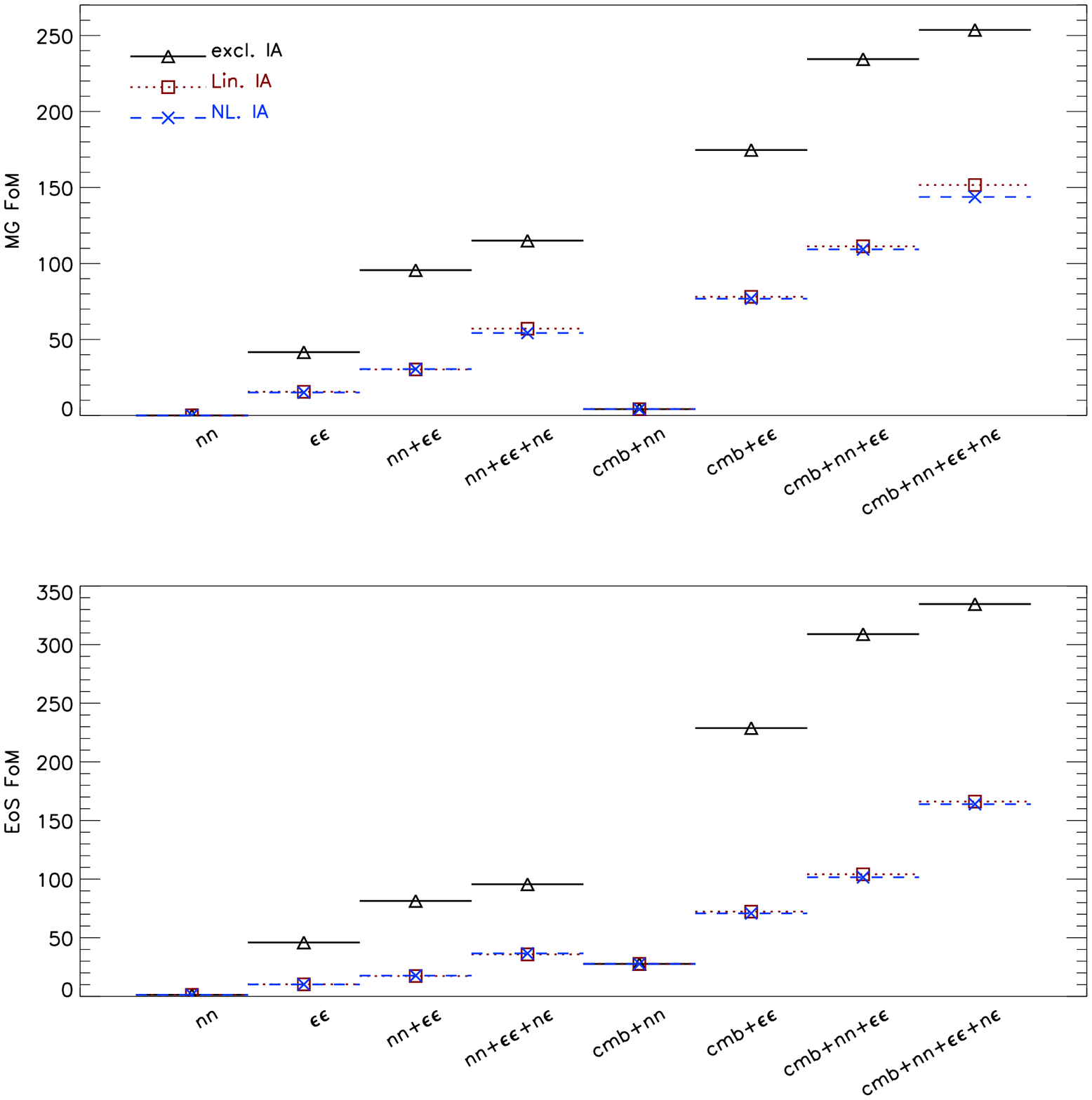}
\caption {
 How figures of merit (FoM)  are affected by intrinsic alignments and the choice of data sets utilised in the analysis for equations of state parameters ($w_0$ vs $w_a$) and modified gravity parameters ($Q_0$ vs. $Q_0(1+R_0)/2$). We compare analyses in which IAs are ignored [black full, triangle], where they are included using the linear alignment model [red dotted lines, square] and the non-linear alignment model [blue dashed, cross]. Datasets include a Planck-like CMB survey, denoted `cmb', and Stage IV galaxy position-position `nn', shear-shear `$\epsilon\epsilon$'  and shear-galaxy position cross-correlations`$n\epsilon$'.}
\label{fig:addindata}
\end{figure*}

Both relaxing the assumption that gravity is described by GR on cosmic scales and adding in systematic uncertainties increase the degrees of freedom fit by the prospective data and hence can degrade the quality of the cosmological information obtained.
In figure \ref{fig:2D} we show these dual effects on the 2D marginalised constraints for the dark energy parameters. The constraints shown are for  all data combined: CMB, galaxy position, lensing shear and cross-correlations, with the conservative bias marginalisation model using an $N_k=N_z=5$ grid.
 Either including IAs, or allowing a modification to  gravity, separately has a roughly comparable effect on reducing the constraining power on $w_0$ and $w_a$, with IAs having a slightly larger impact. The inclusion of intrinsic alignments, while not significantly changing the degeneracy direction, does noticeably weaken the modified gravity parameter constraints.

We quantify the constraining power of the surveys using the covariance matrix for the parameters , $C_{ij}=F_{ij}^{-1}$. The diagonal elements of the covariance matrix give the 1-$\sigma$ measurement uncertainty in each parameter, $\sigma_i = \sqrt{C_{ii}}$. A $2\times2$ submatrix of a pair of parameters, $\tilde{C}(p_i,p_j)$, then gives the figure of merit (FoM) that includes the covariances between the parameters,
\bea
FoM(p_i,p_j) \equiv det[\tilde{C}(p_i,p_j)]^{-1/2}.
\eea
With this definition $FoM=1/\sigma_{eff}^2$, where $\sigma_{eff}$ is the geometric mean of the principal axes of the 2-dimensional error ellipsoid.  Note this differs by a factor of $1/6.17\pi$, from another commonly quoted FoM, the area of the 95\% confidence ellipsoid in the 2D marginalised space.

We consider the Dark Energy Task Force figure of merit on the equation of state (EoS), $FoM(w_0,w_a)$ both in the absence of modifications to gravity, and also when the modified gravity  parameters and included and marginalised over.  We quantify constraints on the modified gravity parameters by considering an equivalent $2\times 2$ FoM, $FoM(Q_0,Q_0(1+R_0)/2) $.

An alternative way to measure dark energy constraints is to consider the combined constraints from equation of state and modified gravity together, through considering the FoM from a $4\times 4$ covariance submatrix,
\bea
FoM(comb)\equiv det[\tilde{C}(w_0,w_a,Q_0,Q_0(1+R_0)/2)]^{-1/4}.
\eea
\cite{Wang:2010gq} applied an analogous statistic to compare constraints on a multi-parameter, model-independent dark energy figure of merit. This FoM accounts for all covariances between the equation of state and modified gravity parameters, but at the same time entangles the EoS and MG constraints, with their different dependencies on measurements of the expansion history and growth of structure.  With this definition, for $n$ parameters, our $FoM$ gives a measure of $1/{\sigma_{eff}}^2$, the mean error in the $n$-dimensional confidence ellipsoid. We note that \cite{Wang:2010gq} define a slightly different $FoM=1/{\sigma_{eff}}^n$, giving a figure of merit that scales with the volume of the $n$-dimensional space.

For a $2\times2$ covariance matrix, $M$, for parameters $x$ and $y$, one can calculate a correlation coefficient $r_{corr} = \sigma_{xy}/\sigma_x\sigma_y=\sqrt{1-det(M)/\sigma_x^2\sigma_y^2}$. By dividing the $4\times4$ covariance matrix for all 4 parameters, $\{w_0,w_a,Q_0,Q_0(1+R_0)/2\}$, into $2\times2$ submatrices, we can define an equivalent  correlation coefficient between the equation of state parameters and modified gravity parameters,
\bea
|r_{corr}| &\equiv & \sqrt{1-\frac{FoM(w_0,w_a)FoM(Q_0,\frac{Q_0(1+R_0)}{2})}{FoM(comb)^2}}. \hspace{0.75cm}
\eea

Table \ref{table-results} summarises the 1-$\sigma$, figure of merit (FoM) and correlation coefficient, $r_{corr}$ results for constraints coming from prospective Stage III and Stage IV surveys. We compare the different IA treatments, and the effect of lensing calibration and photometric redshift offsets.

In the absence of modifications to gravity, IAs still have a significant impact, as was, for example, pointed out in \citep{Joachimi:2009ez}. If IAs are assumed to be perfectly understood then one can achieve a $\sim 14$ fold improvement in the dark energy FoM from the Stage IV survey relative to Stage III. However when astrophysical uncertainties about IAs are included, and marginalised over, we find that the relative improvement of the photometric Stage IV survey, is reduced to 9.

When the modification to gravity described in \ref{cosmomod} is included, measurements of the growth of structure no longer purely constrain $w_0$ and $w_a$. With IAs excluded, the EoS figure of merit with modified gravity allowed is weakened by 50\% relative to GR for a stage III survey, and by 70\% for stage IV.

The inclusion of IA uncertainties reduces both the EoS and MG figures of merit by roughly a factor of 2 relative to those when IAs are excluded from the analysis.  Both with and without IAs included, the modified gravity FoM for a Stage IV survey is roughly a factor 6 improvement over that for Stage III.

Overall, when both modified gravity and dark energy parameters are considered together, the FoM improves by a factor 8.5 between Stage III and Stage IV in the absence of IAs, and this is reduced to just under 7 with the conservative modeling of IAs. The similarities in the adjustments in DE and MG FoM between Stage III and Stage IV suggest a high degree of correlation between the  two parameter pairs. This is quantitatively reinforced by the correlation coefficient $r_{corr}$; the correlations are higher for Stage IV than for Stage III and are degraded by $\sim$15 and 25\%, respectively, with the inclusion of intrinsic alignments.

There is only a small difference between the figures of merit for the linear alignment and nonlinear alignment models are included; with the NLA model giving slightly poorer constraints. The small difference suggests that  the differences between the LA and NLA models are to a large extent drowned out by the uncertainities in the IA bias model. Small scale galaxy position correlations, which would be sensitive to differences in LA vs NLA through the $gI$ term, are typically excluded since the multipoles exceed $\ell_{max}$.

For the analysis shown in the table alone, we also consider the impact of  two additional instrumental systematics: photometric redshift offsets  and lensing shear calibration offsets
, on the figures of merit. Photometric redshift offsets, $\Delta z_i$ alter the galaxy distribution inferred from observations as in (\ref{eq:niz}). When systematic offsets are considered, we model them following \cite{Albrecht:2009ct}: we allow independent offsets in each photometric redshift bins and impose a prior on these offsets of $\sigma(\Delta z_i) = 0.002$. We model shear calibration offsets by altering the measured shear correlations
\bea C_{\ell}^{\epsilon_i\epsilon_j,offset} &=& (1+\Delta m_i)(1+\Delta m_j)C_{\ell}^{\epsilon_i\epsilon_j}
\\
 C_{\ell}^{n_i\epsilon_j,offset} &=&(1+\Delta m_j)C_{\ell}^{n_i\epsilon_j}
\eea
and impose a prior of $\sigma(\Delta m_i) = 0.001\sqrt{N_{ph}}$ in each bin. Shear and redshift calibration offsets introduce an additional 10 parameters.  Shear calibration offsets are qualitatively degenerate with the inclusion of IA correlations of unknown amplitude in the $\epsilon\epsilon$ correlation, and both cause a reduction in the figures of merit.  While the systematic shear offsets in $\epsilon\epsilon$ and $n\epsilon$ are wholly correlated, the IA contributions can differ through the inclusion of the $r_I$ degree of freedom in the $GI$ cross-correlations. As such, the degradation in the constraints from including the instrumental systematic offsets, as we model them here, are not as severe as those from marginalising over the uncertainties in the IA model.

In figure \ref{fig:addindata}, we breakdown the impact of including  IAs on the figures of merit derived as one combines CMB plus galaxy, lensing and galaxy lensing cross correlations in a piece-wise fashion. The inclusion of intrinsic alignments significantly deteriorates the expected dark energy constraints coming from weak lensing on its own, while the use of a grid bias model leads to the galaxy-galaxy correlations providing little constraining power on the non-bias parameters in the model. When lensing and galaxy position data are added in tandem, however,   they are able to provide improved constraints, over and above the lensing data alone. When IAs are included, the combined constraints are noticeably weaker than the constraints predicted by pure shear-shear measurements when IAs are neglected.  The effects are mitigated to a good degree by gaining complementary   information about the underlying cosmological potentials, and isolating out the IAs, by adding in galaxy position data. In particular, the inclusion of cross-correlations, between galaxies and lensing, allow the correlated effects of the $II$ and $GI$ IA contributions to reduce the uncertainties in the IA model.
\begin{figure*}
\center
\includegraphics[scale=0.6]{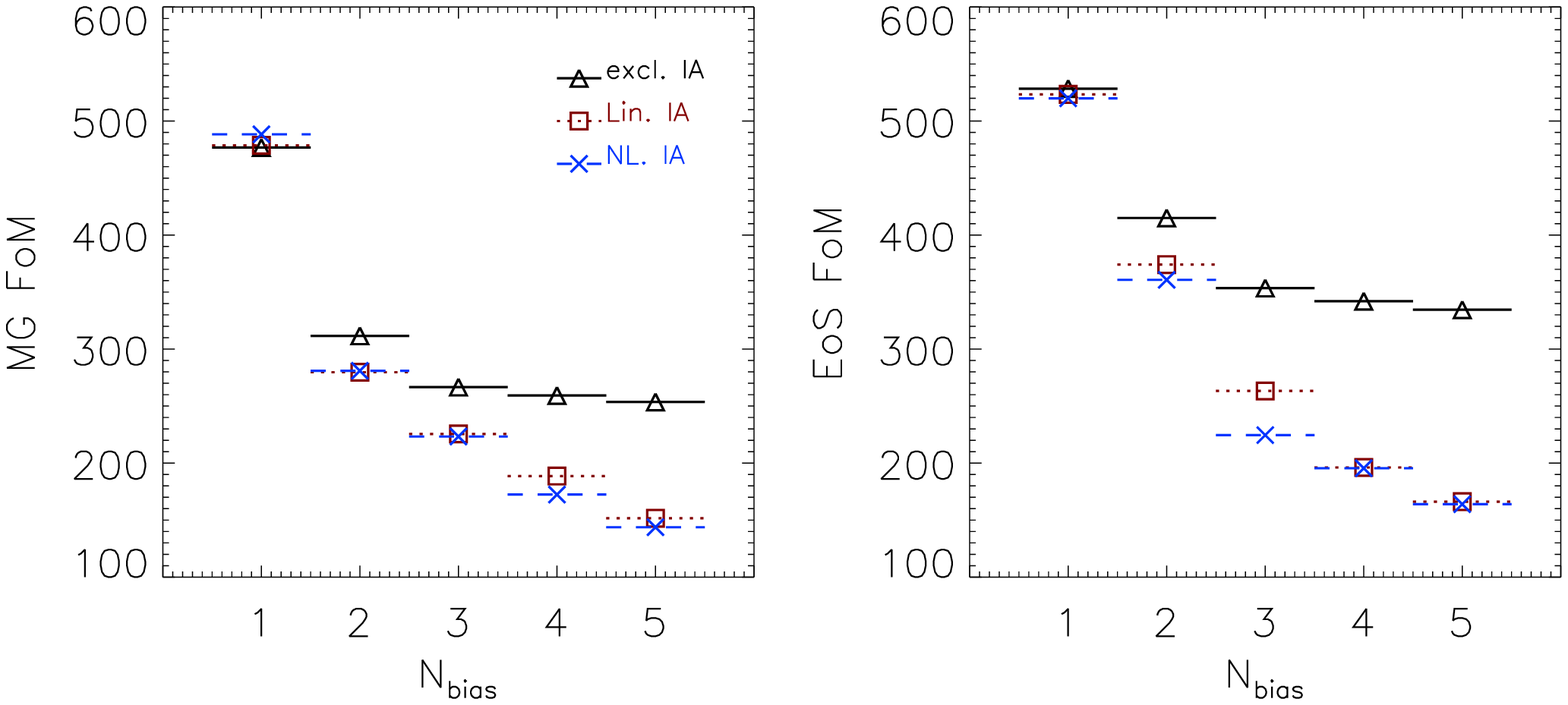}
\caption {
The impact of the number of $k$ and $z$ bins, $N_k=N_z=N_{bias}$, in the bias model on the equation of state (EoS) and modified gravity (MG) figures of merit (FoM). Scenarios are shown in which IAs are excluded [black ,triangle], and in which linear alignment (LA) [red, square] and nonlinear alignment (NLA) [blue, cross] models for intrinsic alignments are used. If IAs are excluded one sees a plateauing of the figure of merit as the number of bias marginalisation parameters in increased. With the addition of parameters to describe uncertainties in the IA amplitude no such plateauing is seen. The inclusion of IA, with an assumption that they are well understood, and can be described by  scale and redshift independent nuisance parameters ($N_{bias}=1$) actually improves the dark energy constraints because the IAs provide additional cosmological information about the high redshift potential $\phi(z_f)$. If uncertainties in the IA model are allowed however, there is a significant deterioration in the constraints on both FoM.
The results presented here are for prospective CMB and Stage IV large scale structure survey utilising all galaxy position and shear auto- and cross- correlations.}
\label{fig:grid}
\end{figure*}
\begin{figure*}
\center
\includegraphics[scale=0.6]{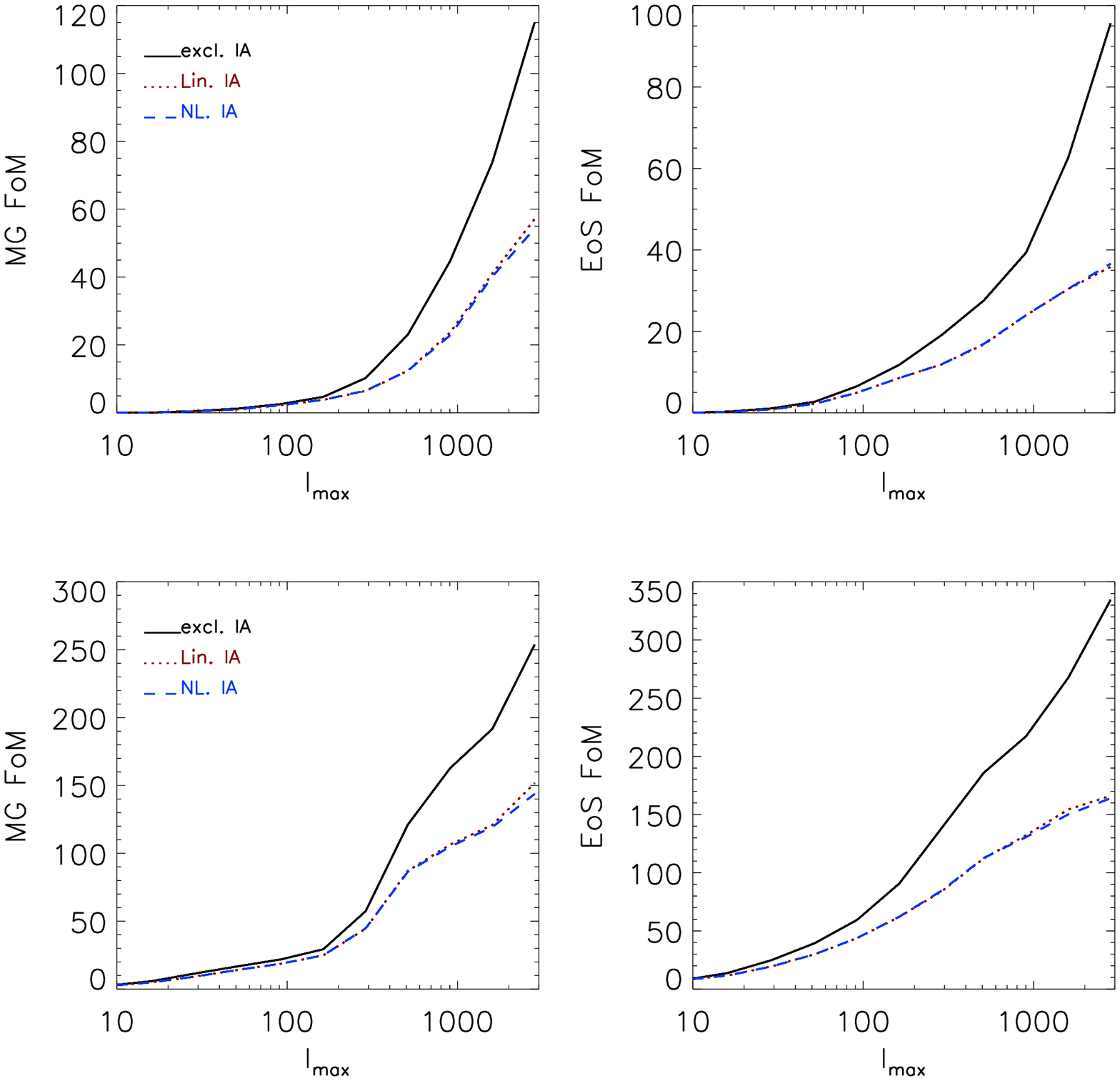}
\caption {
The impact of including observations on small scales, denoted by the maximum multipole, $l_{max}$, up to which correlations are considered, on the equation of state (EoS) and modified gravity (MG) figures of merit (FoM) . Results are shown for a Stage IV photometric survey alone [upper panel] and [lower panel] including complementary constraints from a Planck-like CMB survey when IAs are excluded [black full line] and included using the LA [red,dotted] and NLA [blue,dashed] models. While including smaller-scale observations would appear to improve both figures of merit, one has to consider the theoretical uncertainties present in modeling these small scales, especially in the context of modifications to gravity, therefore it is worthwhile assessing how a conservative approach of neglecting such scales might impact the projected cosmological constraints. }
\label{fig:lsteps}
\end{figure*}

Our findings for  constraints on equation of state parameters are consistent with those of \cite{Joachimi:2009ez}; in the absence of CMB data, figures of merit with all galaxy position and weak lensing correlations and IAs included are comparable with those predicted by weak lensing alone in the absence of IAs.  We do find, however that when we include CMB data  the FoM with IAs never becomes comparable with those when IAs are excluded, even when all cross-correlation information is included.

In a number of recent analyses of dark energy constraints from prospective surveys, the uncertainties in the galaxy bias model are treated by a single, scale and redshift independent, factor. In figure \ref{fig:grid} we highlight  that this assumption can have a dramatic effect on the predicted constraining power of the survey. Allowing a gridded galaxy bias model, while excluding IA uncertainties, reduces the EoS figure of merit by roughly a third, and MG figure of merit by almost a half.

Including intrinsic alignments in the analysis, while assuming single, scale and redshift independent, amplitude has a marginal impact on the EoS and MG FoMs. In fact, interestingly, assuming that you know how IAs are formed and evolve provides additional information, actually improving the constraints.

One can understand this by noting that the GG, II and GI components all depend on the underlying matter distribution but each exhibit a different evolution with redshift. Assuming a single-parameter normalisation, but multiply-binned measurements (from tomography) for the IAs, enables us to obtain an independent measurement of $\psi(z_f)$ from the GG and both IAs.
When uncertainties in the IA model are introduced, however, by using the grid bias model,  they markedly degrade the dark energy constraints. One doesn't see the plateauing of the FoM that one would see with intrinsic alignments excluded from the analysis.

We rationalise the plateau with no intrinsic alignments as follows: the contributions from shear-shear, shear-position and position-position power spectra have different dependencies on redshift resolution. We expect the constraining power of position-position alone to be weak given the multipole cuts, and to be badly affected by even a small number of free bias parameters. Therefore we should be dominated by shear-shear and shear-position information which are much more resilient to a lack of redshift information because, when IAs are not included, each is modulated by the broad lensing weight function.

We can combine the information in figures \ref{fig:addindata} and \ref{fig:grid} to identify how many free parameters can be accommodated before constraints degrade relative to conventional constraints: from shear-shear correlations alone, in which IAs are ignored.
CMB plus shear-shear data alone give a modified gravity figure of merit of 170 (figure \ref{fig:addindata}), which is roughly the same value obtained when including all two-point cross-correlations and including a bias grid with $N_{\rm bias}=4$ (figure \ref{fig:grid}). For the equation of state figure of merit we can use a bias grid with $N_{\rm bias}=3$ before we reach the same figure of merit using all two-point functions as we would obtain from the traditional approach. A bias grid with $N_{\rm bias}=3$ has $3\times3\times4=36$ free parameters in total, $18$ for galaxy bias and a further $18$ for the intrinsic alignment model. We therefore need astrophysics to be sufficiently kind that the bias functions are sufficiently smooth in both scale and redshift, or to have sufficient information from simulations to be able to parameterise the functions with roughly this number of free parameters.

The inclusion of information in the mildly nonlinear regime can have a potentially significant effect on improving constraints on dark energy parameters, purely as a result of the large number of modes available to include in the analysis. If modifications to gravity are included however then one has to make an assessment of how well large scale structure growth in the non-linear regime is understood.
Recent analyses show that in some specific modified gravity theories there can be  subtleties in the non-linear behaviour that might have to be included \citep{Oyaizu:2008sr,Oyaizu:2008tb,Schmidt:2008tn,Khoury:2009tk,Ferraro:2010gh,Cui:2010wb,Zhao:2010qy,Brax:2011ja}. In figure \ref{fig:lsteps}  we highlight the sensitivity of the figure of merit to  the assumptions about the smallest scales to be included in the analysis, parameterised here by $l_{max}$.

When CMB data is included, the pressure to go to high multipoles is reduced. There is only a 50\% increase in FoM on increasing the maximum multipole from 1000 to 3000, compared to over a factor of two when CMB data is not used.

\section{Conclusions}
\label{conclusions}

Using tests of the expansion history of spacetime and the growth of large scale structure, in tandem, gives the best prospects for  testing gravity on cosmic scales. Weak lensing, galaxy position, CMB ISW and potential peculiar velocity observations provide very complementary constraints on the gravitational potentials, through measuring both their sum, $\phi+\psi$ and $\psi$ on its own.

 Fundamental to realising the full potential of these complementary observations is a requirement to minimise both instrumental and astrophysical systematic uncertainties that can dilute the cosmological constraining power of upcoming surveys. Weak lensing observations could potentially offer a direct way to measure the gravitational potentials without the bias uncertainty in relating galaxy positions to the underlying CDM matter distribution.
 On the other hand, intrinsic alignments provide a significant systematic signal.  Uncertainties about IA formation, evolution, and variation amongst galaxy-type, have to be factored into a  realistic assessment of how well weak lensing shear measurements can constrain a cosmological model.

In this paper, we have shown that  how systematic uncertainties are modelled can  have a profound impact on the predicted dark energy  and modified gravity constraints from future large scale structure imaging surveys.

By utilising a  grid-based approach to marginalise over uncertainties in both galaxy bias and intrinsic alignment contributions to lensing shear and galaxy position correlations,  we have provided conservative and optimistic bounds for constraints on both dark energy equation of state parameters and a useful phenomenological modified gravity model.

We considered three figures of merit to quantitatively compare constraints on the dark energy equation of  state and modified gravity parameters both separately and in combination. Quoting separate figures of merit for EoS and MG parameters can be used to show their different dependences on data sets and assumptions. They can also contrast the equation of state parameter dependence on expansion history measurements, and modified gravity parameter dependence on the growth history. We have found, however, that there is a high degree of correlation between these two sets of parameters so that treating them independently ignores an important association. We proposed, and quantified, a combined figure of merit and related correlation coefficient as a way to address this.

We have found that  the constraints have a significant sensitivity to  how galaxy bias and intrinsic alignments are incorporated into the analysis. The  equation of state and modified gravity figures of merit are a factor of 4 smaller  when a conservative scale and redshift dependent grid model is used, than when bias and IAs uncertainities are assumed to be redshift and scale independent. Marginalising over systematic uncertainties  in the IA model  led to a factor of two reduction in the figures of merit.

Whether a linear alignment or nonlinear alignment model underpinned the IA model  had only a small effect in comparison to our assumptions about the evolution of bias and IAs in redshift and scale.  Understanding the astrophysical evolution and population dependence of intrinsic alignments, therefore, could dramatically improve the cosmological information that comes out of future photometric large scale structure surveys, such as DES, SuMIRe, Euclid, LSST and WFIRST. We discuss the implications of weak lensing systematics for optimising cosmic shear surveys to measure dark energy in a related paper \citep{MGPaper2}.

In addition to uncertainties in bias and IAs, an understanding of evolution in the nonlinear regime could also have a profound impact on constraints, through increasing the maximum multipole to which analyses can proceed. If GR governs cosmic evolution this may be achievable, while the model dependence of the non-linear regime in modified gravity models could make this far more challenging.

Combining information from  the photometric surveys we have considered here with that from spectroscopic galaxy data, such as might come from BigBOSS, EUCLID and WFIRST,  might allow closely situated galaxies to be isolated, and their intrinsic alignments to be studied. We will consider in future work how this could provide an important avenue to improve our understanding of intrinsic alignments and in turn maximise the cosmological constraining power of future wide and deep photometric surveys.

We finally note that, as part of this work, we have provided a fitting function to allow other researchers to generate weak lensing and galaxy position correlations for modified gravity theories of the form we consider here.

\section*{Acknowledgements}

We thank the Aspen Center for Physics for support and for hosting two coincident workshops on ``Wide-Fast-Deep Surveys: New Astrophysics Frontier" and  ``Testing General Relativity in the Cosmos" in 2009 where this work was conceived. The authors would like to thank the Kavli Royal Society International Centre for hosting the ``Testing general relativity with cosmology" workshop in 2011 that supported fruitful discussion and collaboration.

We thank Filipe Abdalla, Adam Amara,   David Bacon, Scott Dodelson, Ole Host,  Martin Kilbinger, Andrew Jaffe,  Bhuvnesh Jain, Benjamin Joachimi, Ofer Lahav, Rachel Mandelbaum, Anais Rassat, Alexandre Refregier and Jochen Weller for helpful discussions.

RB's and IL's research is supported by NSF CAREER grant AST0844825, NSF grant PHY0968820, NASA Astrophysics Theory Program grants NNX08AH27G and NNX11AI95G and by Research Corporation.
SLB thanks the Royal Society for support in the form of a University Research Fellowship
and acknowledges support from European Research Council in the form of a Starting Grant with number 240672.

\renewcommand{\thesection}{\Alph{section}}
\setcounter{section}{0}
\section{Appendix: Modified Gravity Fitting Function}
\label{app:fit}

\subsection{Fitting function form}
Some modified gravity (MG) models, such as $f(R)$ theories, can be tailored to reproduce a selected expansion history. However their predictions for the growth of structure can then differ from that predicted by that expansion history assuming GR. Here we consider a fitting function for a modified gravity model in which the expansion history is described by $\Lambda$CDM, but the growth history is modified through a deviation from GR described by two parameters, $Q_0$ and $R_0$, and a third parameter $s$ which encapsulates the time dependence of the deviation:
\bea
 Q(a)&=&1+(Q_0-1)a^s \nonumber
 \\
 R(a)&=&1+(R_0-1)a^s \label{QRsdef}.
\eea
In the main analysis in this paper we have assumed $s=3$.

The key input into calculating the galaxy position and weak lensing shear observables is the matter power spectrum as a function of scale, $k$, and redshift, $z$. Here we obtain a analytical fit for the ratio of the matter power spectrum in the modified gravity model  in (\ref{QRsdef}), to that predicted by  $\Lambda$CDM for the same cosmological parameters:
\bea
r_{fit}(Q_0,R_0,s)=\frac{P(k)_{lin,MG}}{P(k)_{lin,\Lambda CDM}} \label{rfitdef}.
\eea

To motivate the form of the fit, we note that the behaviour of growth in this model is described in \cite{Bean:2010zq} has two distinct regimes as given in (\ref{eq:smallscale}) and (\ref{eq:largescale}). On small scales the growth purely depends on the product $Q_0 R_0$, via (\ref{eq:smallscale}). On large scales, the behaviour involves various derivatives of the modified gravity parameters, and the equations depend uniquely on $Q_0$ and $R_0$. Our fit distinguishes between these two regimes in scale, and fits the evolution with redshift of the high wavenumber (`$H$') and low wavenumber (`$L$') regime separately. The two regimes are joined via a third function, $x(k)$, assuming a transition scale, $k_c$.

\bea
r_{fit}(k,z)& =& \left[1-x(k)\right]r_L(z)+x(k)r_H(z)
\\
r_L(z)& \equiv&1+\left[L_1\left(1-R_0\right)+L_2\left(e^{L_3}-e^{L_3 Q_0}\right)\right]\nonumber \\
&& \times \left(e^{L_4 z}+L_5\right)
\\
\\
r_H(z)& \equiv&1+H_1(1-Q_0 R_0)\left(e^{H_2 z}+H_3\right)
\\
x(k)& \equiv& \tanh\left[\left(\frac{k}{k_c}\right)^p\right]
\eea

The values for the 10 fitting parameters $\{L_1,L_2,L_3,L_4,L_5,H_1,H_2,H_3,k_c,p\}$, are obtained using OriginLab's Origin software to fit our custom function to the power spectra coming from the CAMB code.
While $s=1$ and $s=3$ have been most commonly used choice in the literature, we obtain the fit for $s=1-4$, using a grid of values in $Q_0$ and $R_0$ between 0.9 and 1.1. The spectra were calculated at 50 redshift steps in $0\le z \le3$ and for over a hundred values in k ranging from $7 \times 10^{-6}$  to $40 Mpc^{-1}$.
For $s=3$, 23 different $\{Q_0$,$R_0\}$ pairs were used to obtain the fit, for $s=1$ and $s=2$ a subset of these was used, and found to be sufficient to achieve sub percent accuracy in the $C_\ell$s.

Table \ref{tab:params} provides the values for the fits parameters for each value of  $s$.

\begin{table}
\label{tab:params}
\begin{tabular}{|c ||c |c|c|c|}
\hline
Fit Parameter &$s=1$ & $s=2$ & $s=3$&$s=4$\\
\hline
\multicolumn{1}{|c||}{$L_1$} & 0.5293 & 0.4947 & 0.4268 & 0.3635\\
\hline
\multicolumn{1}{|c||}{$L_2$} & -4.733& -5.692 &  -6.300 & -6.575\\
\hline
\multicolumn{1}{|c||}{$L_3$} & -1.610& -1.660& -1.764 & -1.817\\
\hline
\multicolumn{1}{|c||}{$L_4$} & -0.8678 & -1.610 & -2.409 & -3.263\\
\hline
\multicolumn{1}{|c||}{$L_5$} & 0.2878& 0.0867 & 0.0336 & 0.0156\\
\hline
\multicolumn{1}{|c||}{$H_1$}& -0.5655& -0.2023 & -0.0984 & -0.0557\\
\hline
\multicolumn{1}{|c||}{$H_2$}& -0.6144& -1.263& -1.935 & -2.718\\
\hline
\multicolumn{1}{|c||}{$H_3$}& 0.1803 & 0.0754 & 0.0317 & 0.0153\\
\hline
\multicolumn{1}{|c||}{$k_c [Mpc^{-1}]$}& $1\times10^{-3}$ & $7\times10^{-4}$ & $6\times10^{-4}$ &  $5\times10^{-4}$\\
\hline
\multicolumn{1}{|c||}{p}      & 0.9422 & 1.048 & 1.076 & 1.090\\
\hline
\end{tabular}
\caption{Summary of values for the 10 parameters used in the fitting function, given for each value of $s$, the power law exponent in the modified gravity function (\ref{QRsdef}). }
\end{table}

\subsection{Performance}

The fit given above reproduces the ratios of the matter power spectrum in (\ref{rfitdef}), derived from CAMB, to better than 0.01\% for $k>10 k_c$, within 0.6\% around the transition scale, $k_c$, and within 0.8\% at scales around the horizon scale today. This is sufficient to reproduce the $C_\ell$s in the modified gravity scenario to within sub percent ($\sim$0.1\%) levels. The error is largest for low $\ell$s and increases as $s$ decreases.

\begin{figure}
\center
\includegraphics[scale=0.45]{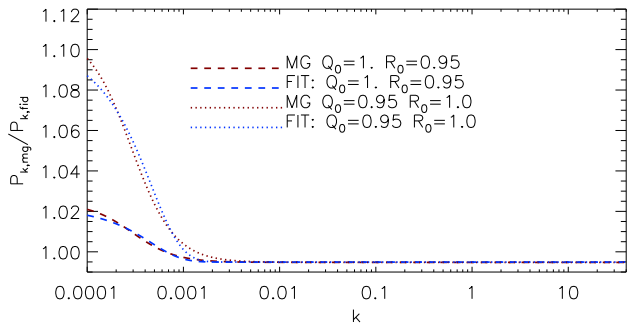}
\caption{
Ratios of $z=0$ matter power spectra in modified gravity model to the fiducial $\Lambda$CDM model obtained via simulation [red] compared with the ratios obtained using the fitting function [blue]. Two models shown are $Q_0=1,R_0=0.95$ [dashed lines] and $Q_0=0.95, R_0=1$ [dotted lines]. At small scales the two models are degenerate, since the evolution of the matter perturbations is only dependent on the product $QR$, while at large scales their behaviours are distinct. The fitting function provides agreement to within 0.01\%  for most scales. At the transition scale $k\sim k_c$ and on horizon scales the fit is a litter poorer, $\sim0.8\%$, however this limited range of scales contributes only a small amount to the angular correlations $C_{\ell}$ used in the analysis. }
\label{fig:pkratios}
\end{figure}

Figure \ref{fig:pkratios} shows the ratio of the matter power spectrum today for the modified gravity model to that for the fiducial $\Lambda$CDM model for both the fit and the full integration using CAMB. At $z=0$, where the modifications are largest, as a function of $k$, the fit matches the simulations to within 0.8 $\%$ and remains accurate at this level for all $z$ for 5\% changes in $Q_0$ and $R_0$.

\begin{figure*}
\center
\includegraphics[scale=0.8]{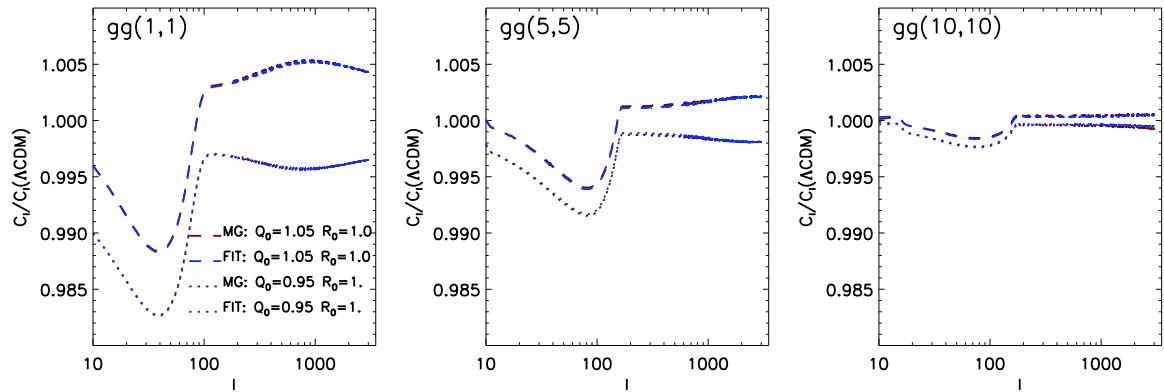}
\caption{
 A comparison of the ratios of the $C_{\ell}$s in the modified gravity model to the fiducial ones obtained via simulation [red] compared to the ratios obtained using the fitting function [blue]. Two models are shown Dashed lines are $Q_0=1.05,R_0=1.00$ [dashed lines] and for $Q_0=0.95, R_0=1$ [dotted lines] . Subpanels left to right indicate correlations in low to high tomographic redshift bins, $1-1$, $5-5$, and $10-10$ respectively. The agreement between fit and simulated $C_{\ell}$s is at the level of $\sim 0.1\%$.}
 \label{fig:CLratios}
\end{figure*}

When calculating the angular correlation function, $C_{\ell}$, the power spectrum fit is integrated over $k,z$.  Applying the fit factor to a standard power spectrum, and using this fit power spectrum to compute $C_{\ell}$s for galaxy autocorrelations, galaxy-weak lensing cross-correlations, and weak lensing auto-correlations results in sub percent($0.1\%$) level accuracy, as shown in figure \ref{fig:CLratios}.

\bibliographystyle{mn2e}
\bibliography{IA}
\label{lastpage}
\end{document}